\documentclass[12pt]{article}
\usepackage{bm}
\usepackage{amsmath}
\usepackage{amssymb}
\usepackage{xcolor}
\usepackage[mathcal]{euscript}
\usepackage{graphicx}
\usepackage{pgf}
\renewcommand{\theequation}{\thesection .\arabic{equation}}

\date{}
\begin{document}
\title{ Determination of the Optimal Elliptical Trajectories Around the Earth and Moon}
\author{Thomas Carter, Professor Emeritus,\\
Department of Mathematics,\\ Eastern Connecticut State University,\\
Willimantic, CT \thanks{e-mail:CARTERT@easternct.edu} \\
and \\
Mayer Humi\\Department of Mathematical Sciences,\\
Worcester Polytechnic Institute,\\
Worcester, MA 01609 \thanks {e-mail: mhumi@wpi.edu}} 

\maketitle
\newpage
\begin{abstract}
Current space exploration programs call for the establishment of a 
permanent Human presence on the Moon. This paper considers periodic orbits 
of a shuttle between the Earth and the Moon. Such a shuttle will be needed to 
bring supplies to the Moon outpost and carry back those resources that are 
in short supply on Earth. To keep this shuttle in permanent periodic orbit 
it must have a thruster that forces it into an elliptical orbit from perigee 
near Earth to an apogee just beyond the Moon and back to perigee. 
The impacts of the Earth, Moon and Sun gravity on this orbit are considered. 
For this model we determine the eccentricity that minimizes the thrust 
requirements and the lunar $\Delta\, v$ requirements. We show that optimal 
placements of the eccentricity of the shuttle orbit can produce significant 
improvement in thrust (and fuel) requirements.

\end{abstract}
\thispagestyle{empty}

%\maketitle

\newpage
\setcounter{equation}{0}
\section{Introduction}

Over the years there have been many suggested trajectories from the vicinity 
of the Earth to the vicinity of the Moon\cite{RFA,BH,SPT,DD,HC,PRP,SV}. 
Extensive research is ongoing also on orbits and trajectories in more 
general contexts.\cite{CM,CD,DHZ,KLM,MTB,MYU,NGZ}
 
The Apollo missions have shown that free-return ballistic figure eight 
shaped trajectories have required only minor orbit adjustments to traverse
between Earth and Moon\cite{SMK}. However such trajectories provide only 
for one round trip between the Earth and the Moon. In view of current plans
to establish permanent human presence on the Moon it is appropriate
to consider a shuttle in a periodic orbit in between the Earth and the Moon. 
In the present paper we consider elliptical orbits for
such a shuttle (This problem was considered briefly in \cite{HC}.). 
We are assuming that the shuttle is equipped with 
a thruster that propels it in a planar elliptical orbit from a perigee 
near Earth to an apogee beyond the moon and back to perigee. The thrust 
negates the out-of-plane effects of solar gravity.
The objective of this paper is to find the eccentricity of the elliptical 
orbit that minimizes the thrust and the lunar $\Delta \,v$ requirements.
It should determine the most efficient of all elliptic orbits for this shuttle
around the Earth and Moon taking into account the Earth Moon and Sun 
gravity on the shuttle orbit.

Complementary studies are assumed. It is likely that
a separate stage will be required for the energy needed to enter the elliptical transfer orbit. For this reason the initial $\Delta\,v$ will constitute a 
separate study not included herein, although we will include a preliminary 
study of the eccentricity effects on the lunar $\Delta\,v$. 
Furthermore we neglect the variation of the Moon orbit from 
circular as well as Earth oblateness and solar radiation of the Sun. 
These effects can be considered in more refined studies or compensated
through a control system.

Relevant standard data involving the Earth, Moon and Sun are available 
in \cite{NAS} 

%%%%%%%%%%%%%%%%%%%%%%%%%%%%%%
\section{The Orbit Requirements}

This preliminary analysis will determine the eccentricity of the orbit which 
optimizes the maximum thrust requirement and also   
the $\Delta v$ requirement needed to traverse between the Moon and the 
spacecraft orbit. This section presents a study of each. The $\Delta \,v$
requirement to traverse from Earth or Earth orbit to the elliptical 
orbit is also highly  important but is not included  in the present study. 
This activity will likely need a separate stage and a separate thruster 
thus requiring another study. 

\subsection{Angular Velocity and Time in Orbit}

Fig $1$ depicts a schematic view of the orbit of a spacecraft 
where the thrust forces a planar elliptical orbit in a frame fixed in the  
axis through
the Earth and the Moon and rotating with the angular velocity $\omega$
of the Moon about the Earth. Relevant points on the Earth-Moon axis
as measured from the Earth's center are as follows: The position of the 
Moon is $x_M$, the perigee of the orbit is $x_p$, the Earth-Moon center 
of gravity is $x_c$, the focus of the elliptical orbit is $x_o$, and 
$x_a$ is the apogee. An arbitrary point on the orbit is denoted by the 
polar pair $(r,\theta)$  as measured from the focus.
It follows that
\begin{equation}
\label{5.1}
r(\theta)=\frac{p}{1+e\cos(\theta)}.
\end{equation}

We denote by $x(\theta)$ the projection of $r(\theta)$ on the Earth-Moon 
axis as measured from the center of the Earth. We observe that 
$r(\theta)\cos(\theta)$ is negative  for $\frac{\pi}{2} < \theta < \pi$.
We can describe $x(\theta)$ as follows:
\begin{equation}
\label{5.2}
x(\theta)=x_0-\frac{p\cos(\theta)}{1+e\cos(\theta)}
\end{equation}
where $x(0)=-x_p$ and $x(\pi)=x_a$. It follows that
\begin{equation}
\label{5.3}
x_p+x_0=\frac{p}{1+e},
\end{equation}
\begin{equation}
\label{5.4}
x_a-x_0=\frac{p}{1-e}.
\end{equation}
Solving these equations for $p$ and $x_0$ we obtain
\begin{equation}
\label{5.5}
p=\frac{x_a+x_p}{2}(1-e^2)
\end{equation}
\begin{equation}
\label{5.6}
x_0=\frac{x_a-x_p}{2}-e\frac{x_a+x_p}{2}.
\end{equation}
\iffalse
For a specified eccentricity the requirements to perform work against gravity  
for half an orbit are
\begin{equation}
\label{5.7}
E(e)=\displaystyle\int_0^{\pi} 
\left[\frac{-g_E}{(x(\theta)^2}+\frac{g_M}{(x_M-x(\theta))^2}\right]
x'(\theta)d\theta 
\end{equation}
where $g_E$ and $g_M$ denote respectively the gravitational constants 
of the Earth and the Moon, and
\begin{equation}
\label{5.8}
x'(\theta)=\frac{p\sin(\theta)}{(1+e\cos(\theta))^2}.
\end{equation}
Evaluating and simplifying the relevant expressions the integral becomes
\begin{equation}
\label{5.9}
E(e)=p\displaystyle\int_0^{\pi}
\left[\frac{-g_E}{(K_E\cos(\theta)-x_0)^2}+
\frac{g_M}{(K_M\cos(\theta)-x_0+x_M)^2}\right]\sin(\theta)d\theta
\end{equation}
\fi
%%%%%%%%%%%%%%%%%%%%%%%%%%%%%%%%%%%%%%%%%%%%%%%%%%%%
%figure 1
\begin{figure}[ht!]
\includegraphics[scale=1,height=160mm,angle=0,width=180mm]{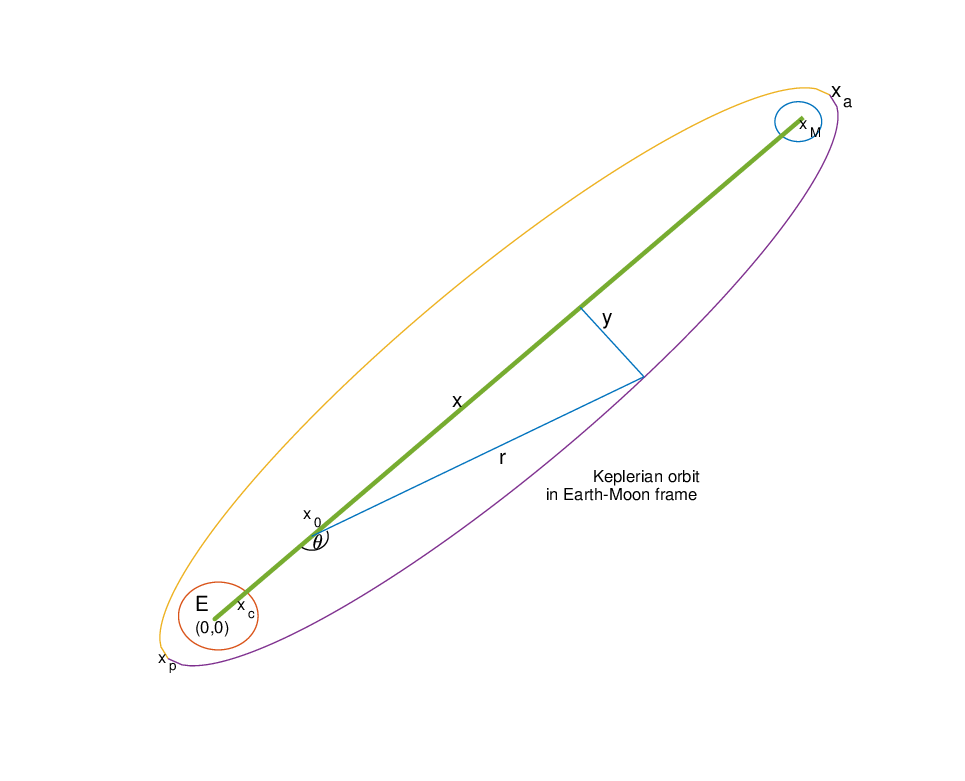}
\caption{A schematic plot of the elliptical orbit about the  Earth and Moon.}
\end{figure}

At the angle $\theta$ the position of the spacecraft (see Figure 1) 
is determined 
by the vector ${\bf r}(\theta)=(x(\theta),y(\theta))$ where $x(\theta)$
is given by \eqref{5.2} and  
\begin{equation}
\label{5.7}
y(\theta)=\frac{p\sin\theta}{1+e\cos\theta}.
\end{equation}
It follows from \eqref{5.2} and \eqref{5.7} that
\begin{equation}
\label{5.8}
{\dot x}(\theta)=\frac{p{\dot \theta}\sin\theta}{(1+e\cos\theta)^2},
\end{equation}
\begin{equation} 
\label{5.9}
{\dot y}(\theta)=\frac{p{\dot \theta}(e+\cos\theta)}{(1+e\cos\theta)^2}.
\end{equation}
The angular velocity function ${\dot \theta}$ is defined as a function of 
$\theta$ on the interval 
$0 \le \theta \le 2\pi$. We may choose this function ${\dot \theta}$ .
The following properties are important
in making this choice of the curve ${\dot\theta}(\theta)$. We require that 
it will be smooth, simple, and symmetric about $\theta=\pi$.
On the basis of these criteria, we select the quadratic function:
\begin{equation}
\label{5.10}
{\dot\theta}(\theta)=-\frac{{\dot\theta}(\pi)-{\dot\theta}(0)}{\pi^2}
(\theta-\pi)^2+{\dot\theta}(\pi).
\end{equation}
We require that ${\dot\theta}(0)\ne 0$ and ${\dot\theta}(\pi)\ne 0$.
It follows that 
\begin{equation}
\label{5.11}
{\ddot\theta}(\theta)=-\frac{2{(\dot\theta}(\pi)-{\dot\theta}(0))}{\pi^2}
(\theta-\pi).
\end{equation}
These expressions are completely determined by ${\dot \theta}(0)$
and ${\dot \theta}(\pi)$. Since ${\dot y}(0)=v_p$ and ${\dot y}(\pi)=v_a$
they can be gotten from \eqref{5.9} and \eqref{5.5}:
\begin{equation} 
\label{5.12}
{\dot\theta}(0)=\frac{(1+e)v_p}{p}=\frac{2v_p}{(1-e)(x_p+x_a)}
\end{equation}
\begin{equation} 
\label{5.13}
{\dot\theta}(\pi)=\frac{(1-e)v_a}{p}=\frac{2v_a}{(1+e)(x_p+x_a)}.
\end{equation}
Time in orbit can be determined from \eqref{5.10} through the integral
\begin{equation} 
\label{5.14}
t(\theta)=\frac{1}{{\dot\theta}(\pi)}
\displaystyle\int_0^{\theta}\frac{ds}{1+\frac{k}{\pi^2}(s-\pi)^2}
\end{equation}
where 
\begin{equation} 
\label{5.14a}
k=\frac{v_p(1+e)}{v_a(1-e)} -1. 
\end{equation}
If $-1 <k <0$ then the integral is evaluated as
\begin{equation} 
\label{5.15}
t(\theta)=\frac{\pi}{2\sqrt{-k}{\dot\theta}(\pi)}
\left[\ln\left| \frac{1+\frac{\sqrt{-k}}{\pi}(\theta-\pi)}
{1-\frac{\sqrt{-k}}{\pi}(\theta-\pi)}\right|+
\ln\left|\frac{1+\sqrt{-k}}{1-\sqrt{-k}}\right|\right],
\end{equation}
if $k=0$, it becomes
\begin{equation} 
\label{5.16}
t(\theta)=\frac{\theta}{{\dot\theta}(\pi)},
\end{equation}
and if $k > 0$
\begin{equation} 
\label{5.17}
t(\theta)=\frac{\pi}{\sqrt{k}{\dot\theta}(\pi)}
\left[Arctan\frac{\sqrt{k}(\theta-\pi)}{\pi}+Arctan(\sqrt{k})\right].
\end{equation}
The time $t_f$ to traverse from perigee to apogee is therefore obtained
from these expressions by setting $\theta=\pi$.

\subsection{Gravitational Effects and Total Thrust}

We transfer from the coordinate system through the Earth and the Moon 
that has been rotated through the angle $\omega t$ as
depicted in Figure 1 to the original system at $t=0$.
Through a coordinate rotation, we obtain
\begin{equation} 
\label{5.18}
X=-y(\theta)\sin\omega t+x(\theta)\cos\omega t
\end{equation}
\begin{equation}       
\label{5.19}
Y=y(\theta)\cos\omega t+x(\theta)\sin\omega t
\end{equation}
where $X$ and $Y$ are measured in the original coordinate system at $t=0$.

We observe that the components of the Earth and 
Moon gravitational forces in the the X-direction respectively are
\begin{equation}
\label{5.20}
F_{E_X}(\theta)=\frac{-g_EX}
{[X^2+Y^2]^{3/2}},
\end{equation}
\begin{equation}
\label{5.21}
F_{M_X}(\theta)=-\frac{g_M(x_M-X)}
{[(x_M-X)^2+Y^2]^{3/2}},
\end{equation}
and in the Y-direction
\begin{equation}
\label{5.22}
F_{E_Y}(\theta)=\frac{-g_EY}
{[X^2+Y^2]^{3/2}},
\end{equation}
\begin{equation}
\label{5.23}
F_{M_Y}(\theta)=\frac{-g_MY}
{[(x_M-X)^2+Y^2]^{3/2}},
\end{equation}
where $g_E$ and $g_M$ denote respectively the gravitational constants 
of the Earth and the Moon.

The gravitational force of the Sun on the spacecraft is written as
\begin{equation}
\label{5.24}
F_S=-\frac{g_S}{r_S^2}
\end{equation}
where $g_S$ is the gravitational constant of the Sun and $r_S$ is the 
distance of the orbital plane of the Earth and Moon from the Sun neglecting
the variation of the distance over the orbit of the spacecraft. The angle 
between the Sun and a normal to the orbital plane is denoted by $\gamma$.
The out-of-plane component of the solar gravitation on the spacecraft is
\begin{equation} 
\label{5.25}
F_{S_Z}=-\frac{g_S\cos\gamma}{r_S^2}.
\end{equation}
%Without thrust compensating for this component, the Shuttle would drift 
%out of the orbital plane by an amount
%\begin{equation} 
%\label{5.26}
%z(t)=-\frac{g_S\cos\gamma}{2r_S^2}t^2
%\end{equation}
%during the time $t$.
The components of the solar gravitation in the X-direction and Y-direction
respectively are
\begin{equation} 
\label{5.27}
F_{S_X}(\theta)=-\frac{g_S\sin\gamma X}
{r_s^2[X^2+Y^2]^{1/2}},
\end{equation}
\begin{equation} 
\label{5.28}
F_{S_Y}(\theta)=-\frac{g_S\sin\gamma Y}
{r_S^2[X^2+Y^2]^{1/2}},
\end{equation}
where $X$ and $Y$ are given respectively by \eqref{5.18} and \eqref{5.19}.

Denoting the total thrust by ${\bf T}$, its components respectively are
\begin{equation} 
\label{5.29}
T_{X}={\ddot X}-F_{E_X}(\theta)-F_{M_X}(\theta)-F_{S_X}(\theta),
\end{equation}
\begin{equation} 
\label{5.30}
T_{Y}={\ddot Y}-F_{E_Y}(\theta)-F_{M_Y}(\theta)-F_{S_Y}(\theta),
\end{equation}
\begin{equation} 
\label{5.31}
T_{Z}=-F_{S_Z}(\theta).
\end{equation}
The expressions ${\ddot X}$ and ${\ddot Y}$ are obtained from \eqref{5.18}
and \eqref{5.19} by differentiation:
\begin{eqnarray} 
\label{5.32}
&&{\ddot X}=-\omega^2[x(\theta)\cos\omega t-y(\theta)\sin\omega t]-
2\omega[{\dot x}(\theta)\sin\omega t+{\dot y}(\theta)\cos\omega t]\\ \notag
&&+{\ddot x}(\theta)\cos\omega t-{\ddot y}(\theta)\sin\omega t
\end{eqnarray}
\begin{equation} 
\label{5.33}
{\ddot Y}=-\omega^2[x(\theta)\sin\omega t+y(\theta)\cos\omega t]-
2\omega({\dot y}(\theta)\sin\omega t-{\dot x(\theta)}\cos\omega t)
+{\ddot x}(\theta)\sin\omega t+{\ddot y}(\theta)\cos\omega t,
\end{equation}
and ${\ddot x}(\theta)$, ${\ddot y}(\theta)$ are the respective derivatives 
of \eqref{5.8} and \eqref{5.9}:
\begin{eqnarray} 
\label{5.34}
{\ddot x}(\theta)=-p\left[\left(\frac{\sin\theta}{(1+e\cos\theta)^2}
-\frac{2e\sin\theta(e+cos\theta)}{(1+e\cos\theta)^3}\right)
{\dot\theta}^2(\theta)- 
\frac{(e+\cos\theta){\ddot\theta}(\theta)}{(1+e\cos\theta)^2}\right]
\end{eqnarray}
\begin{eqnarray} 
\label{5.35}
{\ddot y}(\theta)=-p\left[\left(\frac{\cos\theta}{(1+e\cos\theta)^2}
-\frac{2e\sin^2\theta}{(1+e\cos\theta)^3}\right)
{\dot\theta}^2(\theta)- 
\frac{\sin\theta){\ddot\theta}(\theta)}{(1+e\cos\theta)^2}\right].
\end{eqnarray}
In the Earth-Moon coordinate frame of Figure $1$, the thrust components
respectively are
\begin{equation} 
\label{5.36}
T_{x}(\theta)=-T_Y\sin\omega t+T_X\cos\omega t.
\end{equation}
\begin{equation} 
\label{5.37}
T_{y}(\theta)=T_Y\cos\omega t+T_X\sin\omega t.
\end{equation}
Given $v_p$ we pick various values of $v_a$ and calculate ${\bf T}(\theta)$
from \eqref{5.36}-\eqref{5.37} for values of $\theta$ between $0$ and $2\pi$.
This is done for $0 \le e < 1$ thus finding the value of $e$ that gives
the smallest value of the maximum thrust ${\bf T}(\theta)$ 
for $0 \le \theta \le \pi$.

\subsection{Computational Results}

For $v_p=5.927212$ km/sec and $v_a=7.487672$ km/sec and values of $e$ 
ranging from $0$ to $0.95$ the total thrust was computed in terms of 
$\theta$. For values of $e$ from $0.0$ through $0.5$ the maximum 
thrust could be found at $\theta=0$ as depicted in Figure $2$.
For larger values of $e$ the maximum thrust appears on the interval
$0\le \theta \le \pi$ between $\theta=2$ and $\theta=3$ as demonstrated
in Figures $3-5$. Ranging over the eccentricities  
from $0$ to $0.95$ the smallest value of the maximum thrust was 
$T(0)=1\,\, kg/sec^2$ occurring at $e=0$,
the plot appearing in Figure $6$.

Setting $v_a=v_p=7.487672$ km/sec computations show that a similar situation
occurs. Again the smallest of the maximum thrusts is at $\theta=0$ and the 
smallest of these again at $e=0$.

The simulations were performed again with $v_p=7.409016$ km/sec and 
$v_a=v_p/2$ with similar results. Again the maximum thrust is at $\theta=0$
for eccentricities from $0.0$ through $0.5$ and occurs  between 
$\theta=2$ and $\theta=3$ for eccentricities from $0.7$ to $0.95$.
In all cases the smallest maximum thrust is at $T(0)$. Figure $7$ presents
a plot of the maximum thrust over eccentricity. The smallest value of the
maximum thrust is about $2.35$\,\, $kg/sec^2$. The time to traverse from 
perigee to apogee for this case is calculated from \eqref{5.17} to be about 
$36$ hours.
%fig 02
\newpage
\begin{figure}[ht!]
\includegraphics[scale=1,height=160mm,angle=0,width=180mm]{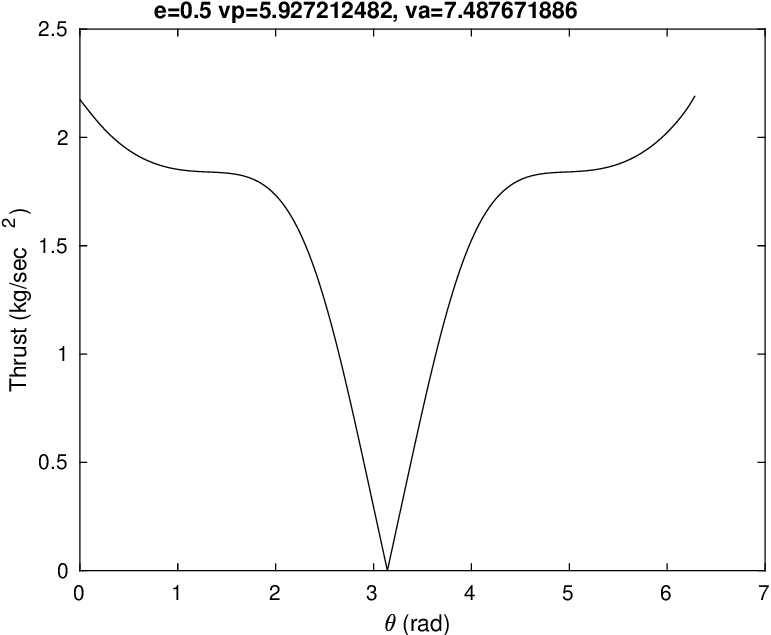}
\caption{Thrust vs. $\theta$,  $e=0.5$, $v_p=5.927212$ $v_a=7.487672$}
\end{figure}
%fig 03
\newpage
\begin{figure}[ht!]
\includegraphics[scale=1,height=160mm,angle=0,width=180mm]{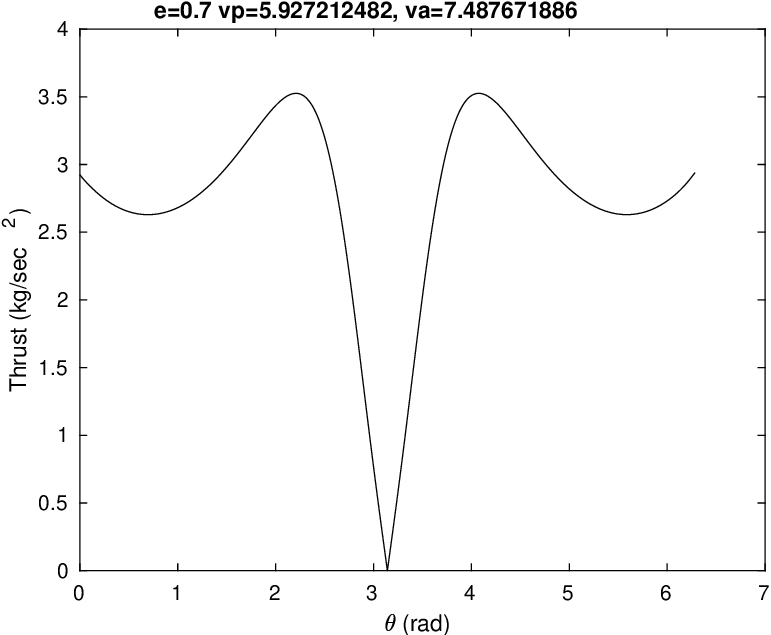}
\caption{Thrust vs. $\theta$,  $e=0.7$, $v_p=5.927212$ $v_a=7.487672$}
\end{figure}
%fig 04
\newpage
\begin{figure}[ht!]
\includegraphics[scale=1,height=160mm,angle=0,width=180mm]{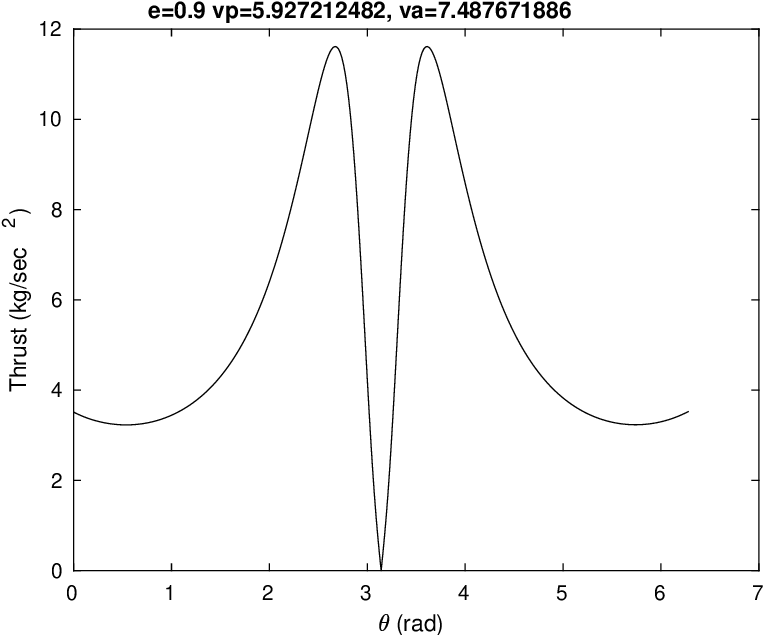}
\caption{Thrust vs. $\theta$,  $e=0.9$, $v_p=5.927212$ $v_a=7.487672$}
\end{figure}
%fig 5
\newpage
\begin{figure}[ht!]
\includegraphics[scale=1,height=160mm,angle=0,width=180mm]{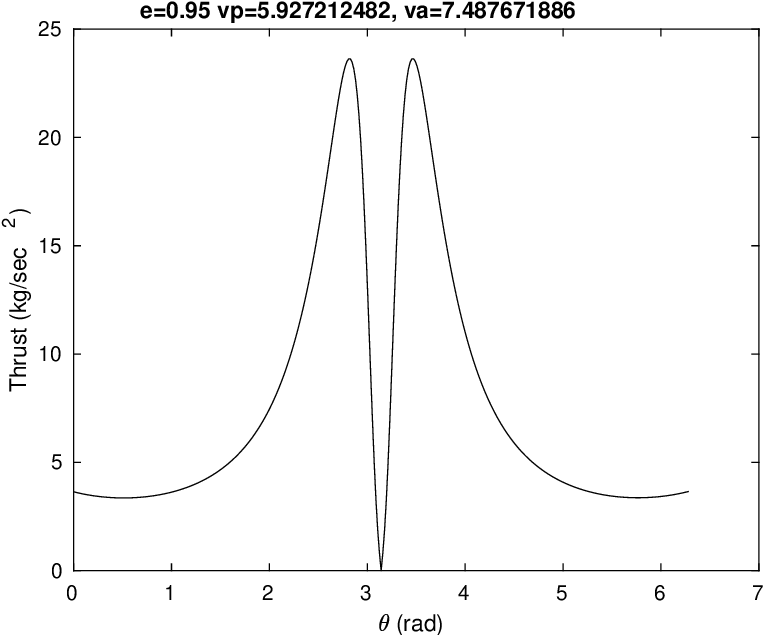}
\caption{Thrust vs. $\theta$,  $e=0.95$, $v_p=5.927212$ $v_a=7.487672$}
\end{figure}

These preliminary studies show that the least thrust requirements are obtained 
from a circular orbit as presented in Figure $8$. There are many thrust 
simulations depending on different values of $v_p$ and $v_a$ that produce 
this same circle but the times $t_f$ in proceeding from perigee to apogee
will vary.
%fig 6
\newpage
\begin{figure}[ht!]
\includegraphics[scale=1,height=160mm,angle=0,width=180mm]{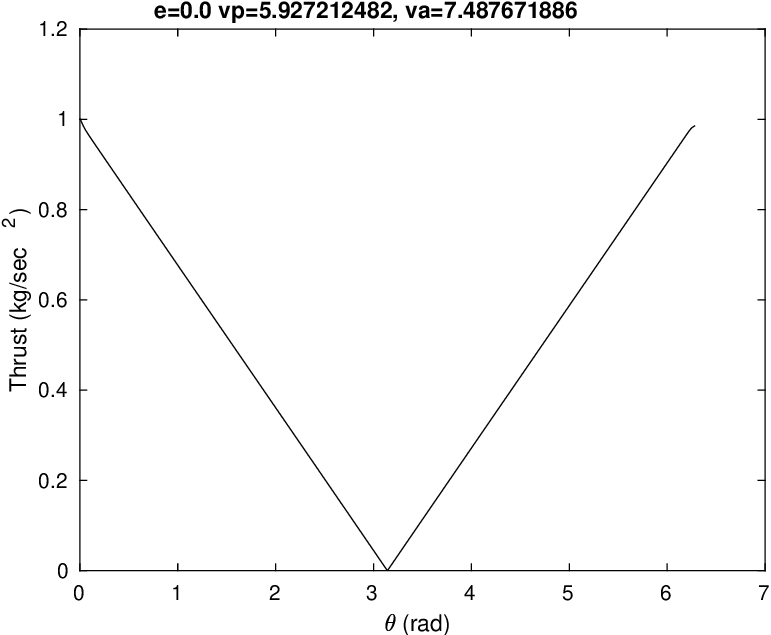}
\caption{Thrust vs. $\theta$,  $e=0$, $v_p=5.927212$ $v_a=7.487672$}
\end{figure}
%fig 7
\newpage
\begin{figure}[ht!]
\includegraphics[scale=1,height=160mm,angle=0,width=180mm]{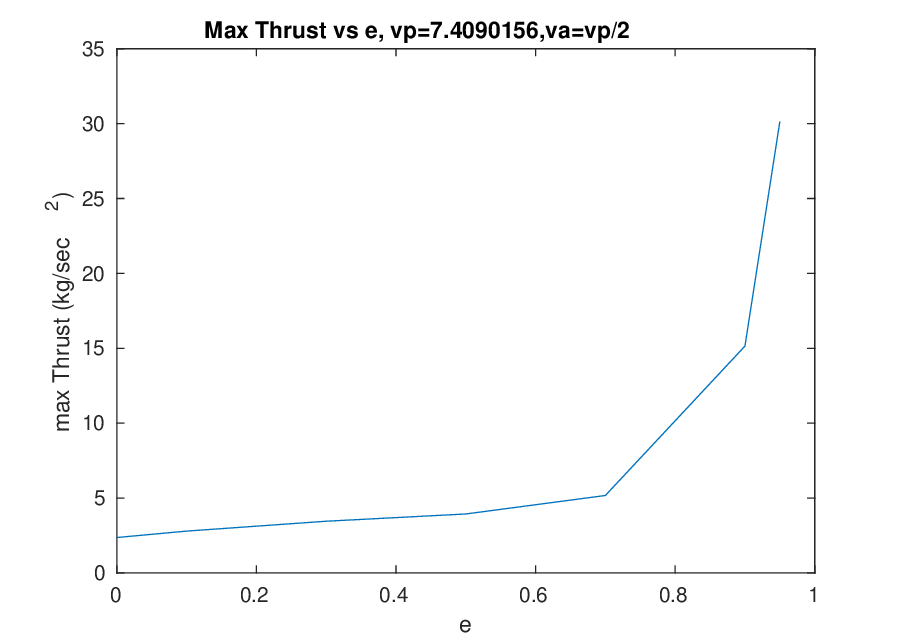}
\caption{Maximum Thrust vs. $e$,\,\,\,$v_p=7.409015$,\, $v_a=v_p/2$}
\end{figure}
{
\subsection{Optimal Velocity at Apogee}

We are also interested in determining the value of the eccentricity $e$
that produces the optimal lunar $\Delta\, v$, the velocity between apogee and 
the lunar orbit or lunar surface.This is equivalent to producing a minimum of 
$v_a$

In this part of  the study we are assuming that $v_p > v_a$, so it follows 
from \eqref{5.14a} that $k> 0$ consequently \eqref{5.17} implies that the total
time from perigee to to apogee is
\begin{equation}
\label{5.38}
t_f=\frac{\pi Arctan{\sqrt{k}}}{{\dot \theta}(\pi)}\sqrt{k}.
\end{equation} 

Solving for ${\dot \theta}(\pi)$ and using \eqref{5.13} we obtain
\begin{equation}
\label{5.39}
v_a=\frac{\pi}{2}(x_p+x_a)(1+e)\frac{Arctan{\sqrt{k}}}{t_f\sqrt{k}}.
\end{equation}
This yields the following nonlinear equation in $k$
\begin{equation}
\label{5.40}
\frac{\pi(x_p+x_a)(1-e)(k+1)}{2v_pt_f}\frac{Arctan{\sqrt{k}}}{\sqrt{k}}=1.
\end{equation}

Given $v_p$ and $t_f$ we solve this equation for  $k$ in terms of $e$.
We can then determine the velocity at apogee $v_a$ from \eqref{5.39}.
For $v_p=5.927$ km/sec and constant $t_f=105000$ sec 
we present a plot of $v_a$ versus $e$ in Figure $9$. 
This plot shows that also for this problem, the minimizing 
eccentricity occurs at $e=0$.}

\section{Conclusions}

{This study calculated the thrust requirements to 
direct an Earth-Moon spacecraft into a periodic elliptical orbit. 
It was found that the least thrust requirements are obtained 
if the orbit consists of a circle through the perigee and apogee. 
The same result was also found to minimize the $\Delta \,v$ between 
this orbit and the lunar orbit or the lunar surface.
It is impressive how much the thrust requirements can be reduced by 
setting the eccentricity of the orbit near zero. It should be possible 
to devise a control system that drives the craft back to the designated 
orbit in order to compensate for disturbances that were not considered 
in the analysis.} 
It might be argued that could one could guess that the circular orbit of the 
shuttle provides the optimal orbit in terms of thrust.  However this 
orbit has a trajectory of maximum length. It follows then that the result 
obtained in this paper though it might be "intuitively clear" is not 
necessarily obvious.

%figure 8
\newpage
\begin{figure}[ht!]
\includegraphics[scale=1,height=160mm,angle=0,width=180mm]{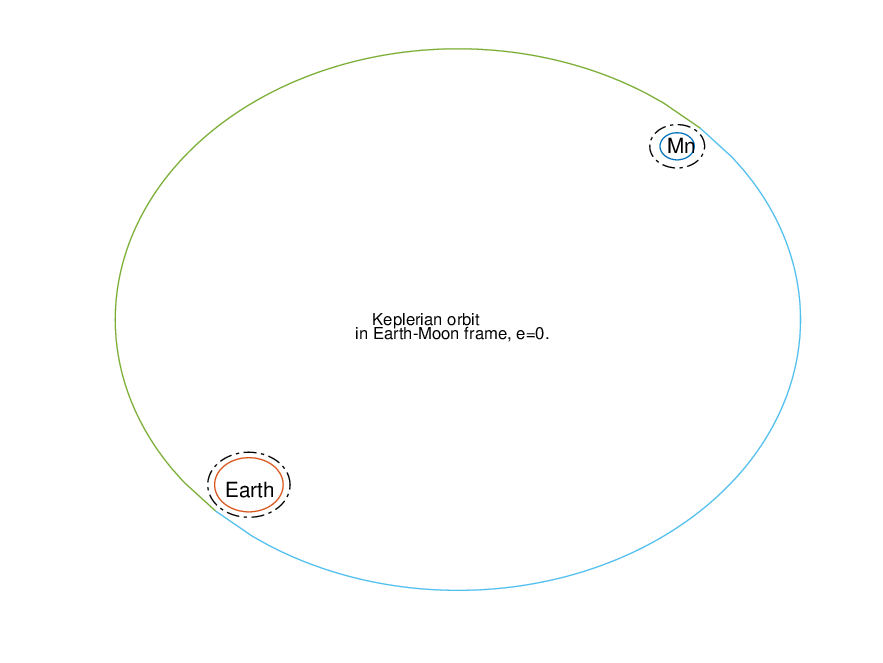}
\caption{A schematic plot for the most efficient elliptical orbit.}
\end{figure}
%figure 9
\newpage
\begin{figure}[ht!]
\includegraphics[scale=1,height=160mm,angle=0,width=180mm]{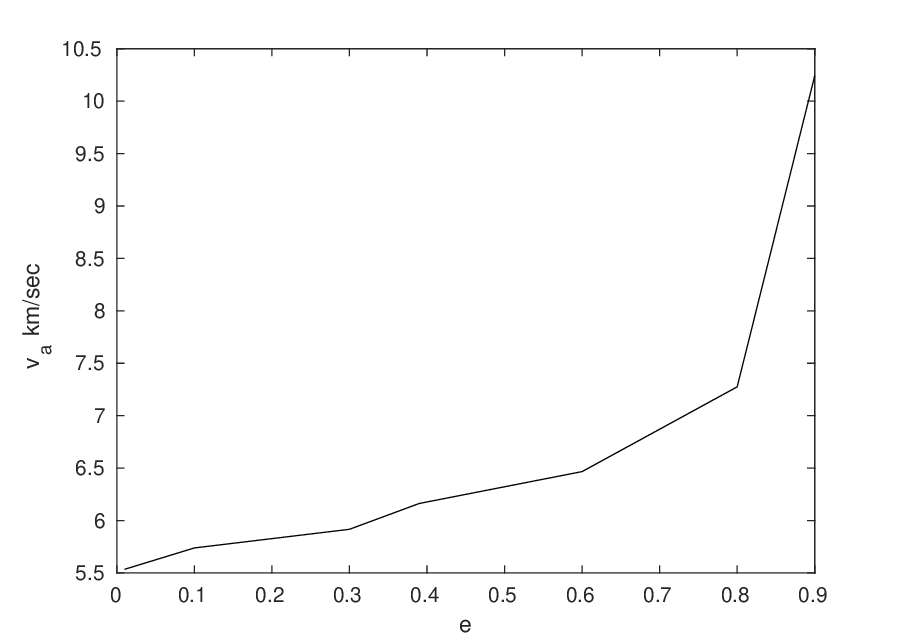}
\caption{Velocity at Apogee vs. $e$: $v_p=5.927$ km/sec, $t_f=105000$ sec.}
\end{figure}

\end{document}